\documentclass[10pt,letterpaper]{article}

\usepackage[margin=1in]{geometry}
\usepackage{amsmath,amssymb,amsfonts}
\usepackage{graphicx}
\usepackage{booktabs}
\usepackage{placeins}
\usepackage{hyperref}
\usepackage{xcolor}
\usepackage{cite}
\usepackage{tikz}
\usetikzlibrary{positioning,arrows.meta}

\hypersetup{
  colorlinks=true,
  linkcolor=blue,
  citecolor=blue,
  urlcolor=blue
}

\title{Quantum Sidecar Architectures for Hybrid AI Training and Inference:\\
Stateful Protected Registers, Stateless Reset-and-Reprepare Circuits,\\
and Quantum Weight-State Outlook}

\author{
Y. Mo\\
Independent Researcher; BroadLink Co., Ltd.\\
\and
G. D. Su\thanks{Corresponding author: \texttt{guodong@hdu.edu.cn}}\\
Associate Professor, Hangzhou Dianzi University
}

\date{\today}

\begin{document}
\maketitle

\begin{abstract}
We propose a quantum sidecar architecture family for future hybrid AI training
and inference. The central idea is not to store an entire Transformer in a
small quantum memory, nor to claim one-shot collapse into a fully trained model
or an optimal answer. Instead, we identify two physically distinct operating
modes for quantum co-processors attached to classical large-model pipelines.
The first is a \emph{stateful protected-register mode}, in which a protected
register stores a reusable quantum resource while an ancilla or temporary
register performs QND-style readout. The second is a \emph{stateless
reset-and-reprepare mode}, in which each query prepares a task-conditioned
quantum circuit, evolves over bounded training or inference control variables,
measures candidate signals, resets the qubits, and repeats. We simulate the
stateful mode using 2/4/6/8 protected-qubit density-matrix QND-style parity
readout with one ancilla and a Qiskit cross-check. For the stateless mode, we include both an abstract
candidate-update sampler and a circuit-level QAOA-style statevector sampler
over structured candidate landscapes, followed by reset-overhead sensitivity
analysis. The resulting framework positions quantum sidecars as bounded signal generators for
optimizer-side sampling, adapter or expert selection, retrieval, routing, and
reasoning-path proposal. As a speculative outlook, we introduce
\emph{quantum weight-state sidecars}: restricted quantum representations over
model-control variables, not direct encodings of complete classical weight
tensors.
\end{abstract}

\section{Introduction}

Large AI models are constrained by training cost, memory movement, routing and
retrieval overhead, and sequential autoregressive inference. Scaling
Transformers to hundreds of billions or trillions of parameters has improved
capabilities, but it also increases optimizer state, activation memory,
key-value cache traffic, and serving cost \cite{vaswani2017attention,
brown2020language,fedus2022switch}. These bottlenecks motivate new computing
interfaces rather than only larger classical accelerators.

Quantum computation offers superposition, unitary evolution, measurement,
interference, and reset as native operations. However, several tempting claims
must be avoided. An $n$-qubit state lives in a $2^n$-dimensional Hilbert space,
but this does not mean that $n$ qubits can store and return $2^n$ independent
classical parameters. Quantum parallelism is useful only when a circuit,
Hamiltonian, or oracle turns amplitudes into a measurable advantage through
interference. Measurement returns limited classical information.

This paper therefore frames quantum computing as a \emph{sidecar} for
classical AI systems. A sidecar does not replace the full model. It proposes,
samples, scores, or filters bounded control variables that the classical model
can consume: update directions, adapter choices, expert gates, retrieval
choices, sampling policies, or reasoning-path proposals. We study two
complementary modes:
\begin{enumerate}
  \item \emph{Stateful protected-register mode}: a reusable quantum resource is
  preserved across queries, while temporary readout registers are measured and
  reset.
  \item \emph{Stateless reset-and-reprepare mode}: no long-lived protected
  resource is assumed; each query prepares, evolves, measures, resets, and
  repeats.
\end{enumerate}

\paragraph{Claim boundary.}
This work does not claim arbitrary state cloning, complete quantum replacement
of trillion-parameter weight tensors, or guaranteed one-shot generation of an
optimal answer. It proposes an architecture family and small simulations that
make the training and inference interfaces concrete.

\begin{figure}[!htbp]
  \centering
  \begin{tikzpicture}[
    font=\small,
    systembox/.style={
      draw,
      rounded corners,
      thick,
      align=center,
      text width=4.8cm,
      minimum height=2.6cm,
      inner sep=5pt,
      fill=blue!4
    },
    sidecarbox/.style={
      draw,
      rounded corners,
      thick,
      align=center,
      text width=3.1cm,
      minimum height=2.1cm,
      inner sep=5pt,
      fill=green!5
    },
    signalbox/.style={
      align=center,
      font=\scriptsize,
      fill=white,
      inner sep=2pt
    },
    signal/.style={-{Latex[length=2.8mm]}, thick}
  ]
    \node[systembox] (llm) at (0,0) {
      \textbf{Large Classical LLM System}\\
      GPU/TPU cluster\\
      Transformer weights\\
      KV cache and optimizer state
    };

    \node[sidecarbox] (sidecar) at (9.6,0) {
      \textbf{Quantum Sidecar}\\
      few-qubit circuit\\
      sampler/readout module
    };

    \draw[signal] ([yshift=0.55cm]llm.east) -- ([yshift=0.55cm]sidecar.west);
    \node[signalbox, text width=3.0cm] at (4.8,0.95) {
      bounded context\\
      candidate set
    };

    \draw[signal] ([yshift=-0.55cm]sidecar.west) -- ([yshift=-0.55cm]llm.east);
    \node[signalbox, text width=3.8cm] at (4.8,-1.05) {
      expert selection\\
      routing hints\\
      gradient/update proposals
    };
  \end{tikzpicture}
  \caption{High-level quantum sidecar interface for LLM systems. The classical
  LLM system keeps the full model state, while the quantum sidecar returns
  bounded control signals such as expert selection, routing hints, and
  candidate gradient/update proposals.}
  \label{fig:sidecar_overview}
\end{figure}
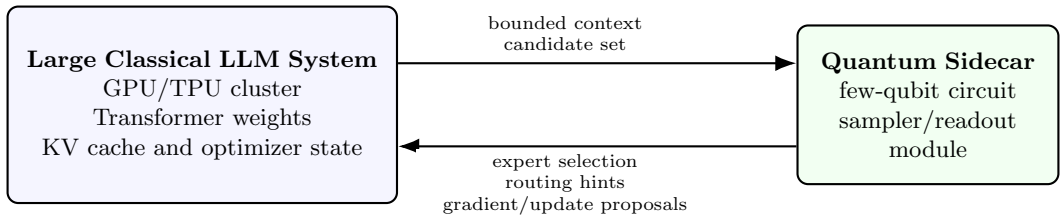

\section{Background}

\subsection{Transformer-scale systems}

Transformer models use attention and feed-forward blocks to map token sequences
to contextual representations \cite{vaswani2017attention}. Large-scale systems
add retrieval, mixture-of-experts routing, long-context decoding, verifier
models, and preference-optimization loops \cite{lewis2020retrieval,
fedus2022switch}. These systems contain many bounded decision points:
selecting experts, adapters, retrieval passages, candidate updates, decoding
paths, or verifier calls. Such decisions are narrower targets for quantum
sidecars than full model replacement.

\subsection{Quantum information constraints}

The no-cloning theorem forbids a universal operation that copies arbitrary
unknown nonorthogonal quantum states \cite{wootters1982single,
dieks1982communication}. The exponential dimension of Hilbert space is not
freely readable classical memory \cite{nielsen2010quantum}. Data loading and
state preparation can dominate any claimed advantage. These constraints shape
the two modes studied here: the stateful mode reads selected observables
without directly measuring the protected register, while the stateless mode
avoids persistent quantum memory and instead re-prepares a circuit for each
query.

\section{Related Work and Positioning}

This work is closest to hybrid quantum-classical machine learning, but its
claim is architectural rather than algorithmic
\cite{biamonte2017quantum,schuld2018supervised}. Quantum kernel methods already
provide a clean sidecar pattern: a quantum device estimates feature-space
overlaps or kernel entries, while a classical learner consumes those quantities
\cite{havlicek2019supervised}. Variational quantum algorithms, including VQE
and QAOA, already use the prepare-evolve-measure-reset loop studied here
\cite{peruzzo2014variational,farhi2014qaoa,cerezo2021variational}. Our
stateless mode should therefore be read as a systems-level use of VQA-like
circuits as AI signal sources, not as a new primitive replacing VQA.

Generative quantum models, including QGANs and quantum circuit Born machines,
also rely on repeated state preparation and measurement to produce samples
\cite{lloyd2018quantum,dallaire2018quantum,benedetti2019generative}. The
sidecar framing differs in where the samples are consumed: the output is not a
standalone generated object, but a bounded control signal for a classical
training or inference pipeline. For training-side use, analytic gradients from
parameter-shift rules and noisy finite-difference methods such as SPSA are
standard mechanisms for updating circuit parameters
\cite{mitarai2018quantum,schuld2019evaluating,spall1992spsa}. Hadamard-test and
overlap-estimation primitives provide related signal-estimation tools, although
they often require deeper or more coherent circuits than the small samplers
simulated in this paper \cite{nielsen2010quantum}.

The intended novelty is therefore not the invention of QAOA, VQE, QGANs,
quantum kernels, or gradient rules. It is the explicit separation of
\emph{stateful} protected-register sidecars from \emph{stateless}
reset-and-reprepare sidecars, together with conservative AI-facing interfaces
that avoid claims of full weight replacement or one-shot optimal inference.

\section{Architecture Family}

\subsection{Stateful protected-register mode}

The stateful mode separates a protected register $A$ from a temporary register
or ancilla $B$. Register $A$ stores a reusable quantum resource. Register $B$
extracts selected observables, is measured, and is reset. The key pattern is:
\[
  A\ \mathrm{stores}, \quad B\ \mathrm{is\ consumed}.
\]
This is inspired by QND measurement, stabilizer readout, and syndrome
extraction \cite{braginsky1996qnd,gottesman1997stabilizer}. It is appropriate
when the same quantum resource must be queried many times and state
preparation is expensive.

\subsection{Stateless reset-and-reprepare mode}

The stateless mode removes the protected register assumption. Each query
executes the standard near-term quantum loop:
\[
  \mathrm{prepare}\rightarrow \mathrm{evolve}\rightarrow
  \mathrm{measure}\rightarrow \mathrm{reset}\rightarrow \mathrm{repeat}.
\]
The quantum sidecar receives bounded classical control variables from a
training or inference loop, prepares a task-conditioned circuit, evolves over a
candidate space, measures one or more candidate signals, and resets all qubits.
This mode is closer to current NISQ hardware and does not require QND storage
or A-to-B mapping. Its central bottlenecks are state preparation, measurement,
shot count, circuit depth, and classical-quantum orchestration, not reset
alone.

\subsection{Training and inference interfaces}

For training, the sidecar should not generate complete model weights. Instead,
it can propose candidate update directions, low-rank adapter updates, prompt
parameters, expert gates, or optimizer perturbations. The classical model keeps
its weights in classical memory:
\[
  W_{t+1} = W_t + \eta \Delta,
\]
where $\Delta$ is selected or proposed by a sidecar and then evaluated by the
classical training loop.

For inference, the sidecar can propose routing, retrieval, sampling, or
reasoning-path signals. The output is a bounded signal, not a full
wavefunction:
\[
  s \in \{\mathrm{expert},\mathrm{passage},\mathrm{path},\mathrm{candidate}\}.
\]

\subsection{Choosing between the two modes}

The two modes are intended for different engineering regimes. The stateful
mode is appropriate when state preparation is expensive, the same quantum
resource can be reused across many calls, and the hardware supports protected
readout of selected observables. Examples include stabilizer-like resources,
quantum random feature sources, or repeated observable probes.

The stateless mode is appropriate when each query carries different context or
candidate variables, when near-term hardware does not support long-lived
protected resources, or when reset and re-preparation are cheaper than
preserving a persistent state. This is the more natural NISQ deployment path:
prepare a task-conditioned circuit, measure a bounded signal, reset all
qubits, and repeat. A practical AI system may combine both modes: stateful
resources for reusable side information and stateless circuits for
query-specific sampling.

\section{Stateful QND-Style Readout Experiment}

\subsection{Parity readout model}

We use an $m$-qubit protected register and one ancilla. The protected state is
a GHZ-like even-parity state:
\[
|\psi_A\rangle =
\alpha |0\ldots0\rangle + e^{i\phi}\beta |1\ldots1\rangle,
\quad |\alpha|^2+|\beta|^2=1.
\]
For even $m$, this state is an eigenstate of the global parity observable
\[
P_A = Z_1 Z_2 \cdots Z_m.
\]
The readout unitary applies controlled-NOT gates from each protected qubit to
the ancilla:
\[
U_{\mathrm{parity}} = \prod_{j=1}^{m}\mathrm{CNOT}(A_j\rightarrow B).
\]
In the ideal case, measuring the ancilla reads parity without revealing which
branch of the superposition is present.

\subsection{Noise model}

We simulate repeated readout with density matrices. Each round applies
$U_{\mathrm{parity}}$, then independent single-qubit depolarizing channels to
each protected qubit and to the ancilla. For one target qubit, the channel is
\[
\mathcal{E}_p(\rho)=(1-p)\rho+\frac{p}{3}(X\rho X+Y\rho Y+Z\rho Z).
\]
The full round applies this channel sequentially across all $m+1$ qubits. We
track fidelity $F_t=\langle\psi_A|\rho_A^{(t)}|\psi_A\rangle$. Since the
simulated protected states have even parity, the parity readout accuracy is
$q_t=\Pr(B=0)$.

\subsection{Results}

\begin{figure}[!htbp]
  \centering
  \includegraphics[width=0.88\linewidth]{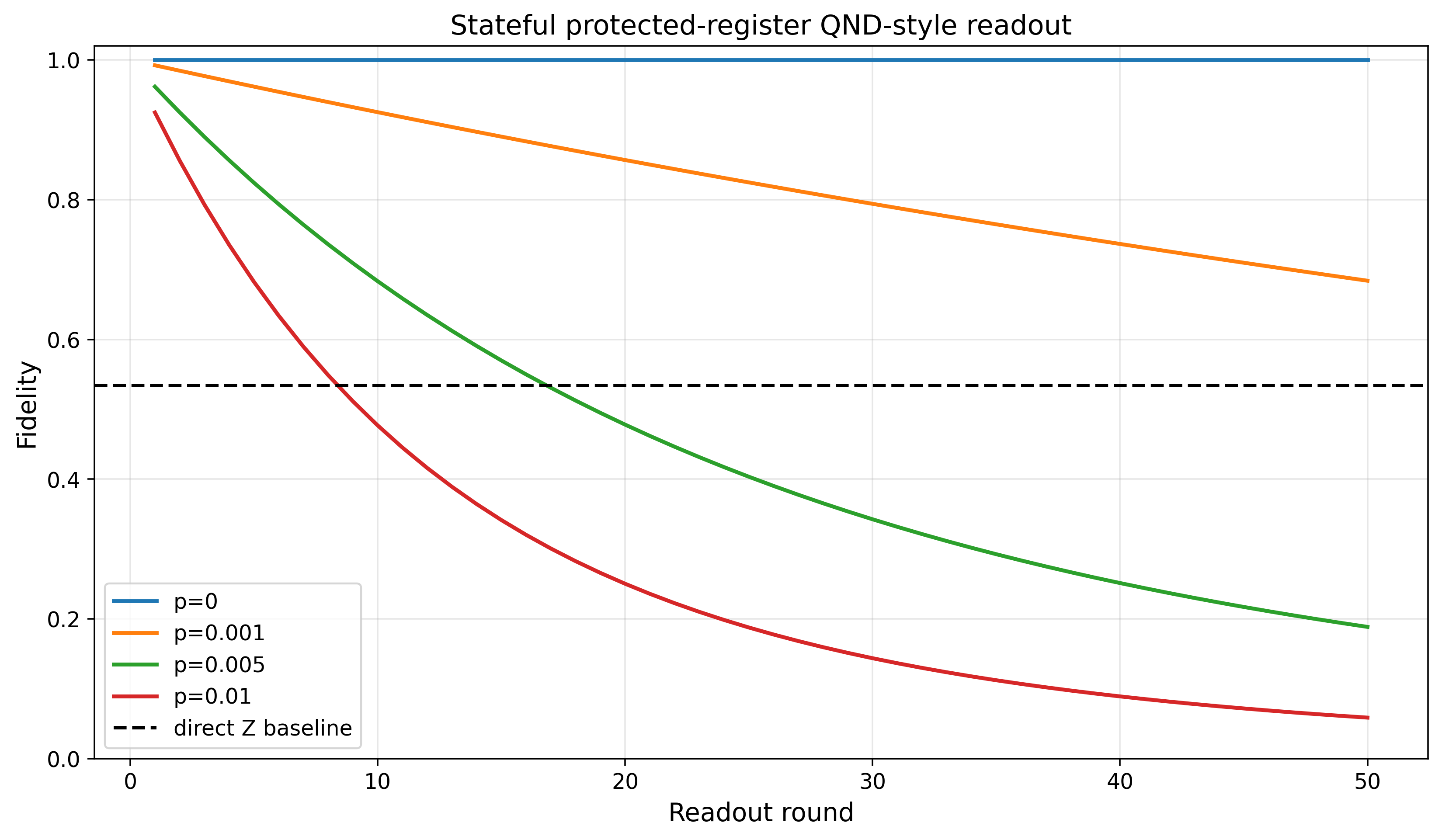}
  \caption{Stateful protected-register QND-style readout for 8 protected qubits
  plus one ancilla. The ideal case is stable; noise causes gradual
  degradation.}
  \label{fig:qnd}
\end{figure}

We set $|\alpha|^2=0.37$, $|\beta|^2=0.63$, and $\phi=0.41$. Fig.~\ref{fig:qnd}
shows repeated 8-qubit readout. The ideal primitive preserves the protected
state, while depolarizing noise accumulates. The direct computational-basis
baseline is $|\alpha|^4+|\beta|^4=0.37^2+0.63^2=0.5338$, because direct readout
removes the off-diagonal coherence between $|0\ldots0\rangle$ and
$|1\ldots1\rangle$.

We cross-check the ideal parity-readout circuit using Qiskit's
\texttt{DensityMatrix}, \texttt{Operator}, and partial-trace utilities. The
Qiskit result matches the NumPy implementation with trace distance $0.0$ and
maximum absolute matrix-entry difference $0.0$ for the checked circuit.

\begin{figure}[!htbp]
  \centering
  \includegraphics[width=0.88\linewidth]{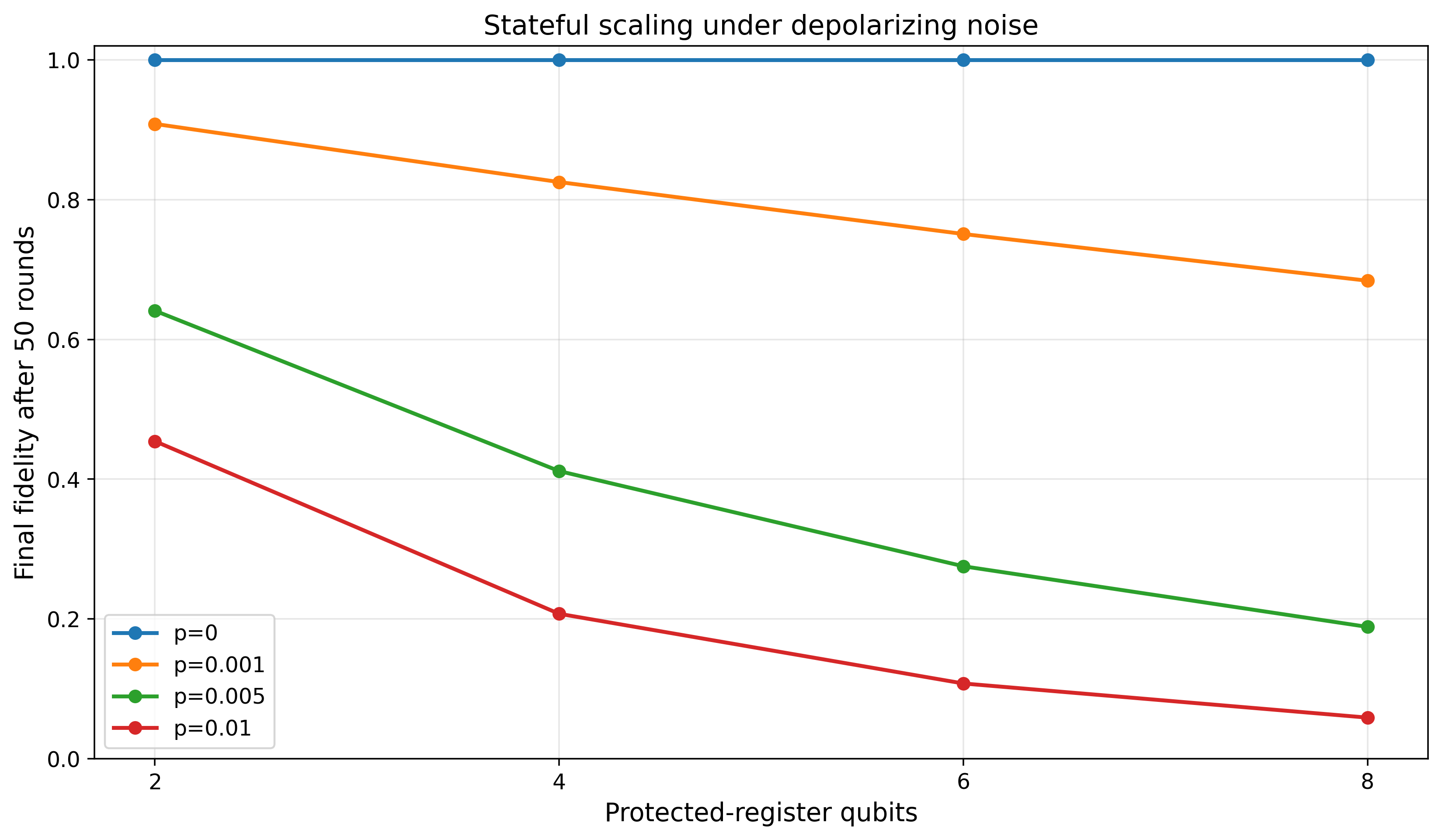}
  \caption{Stateful scaling under depolarizing noise after 50 readout rounds.}
  \label{fig:stateful_scaling}
\end{figure}

Fig.~\ref{fig:stateful_scaling} shows the 2/4/6/8 scaling study after 50
rounds. The noiseless case remains stable for all tested sizes; noisy cases
degrade faster as register size grows because the channel acts on more qubits
per round.

\FloatBarrier

\section{Stateless Reset-and-Reprepare Experiments}

\subsection{Candidate update-direction search}

\begin{figure}[!htbp]
  \centering
  \includegraphics[width=0.88\linewidth]{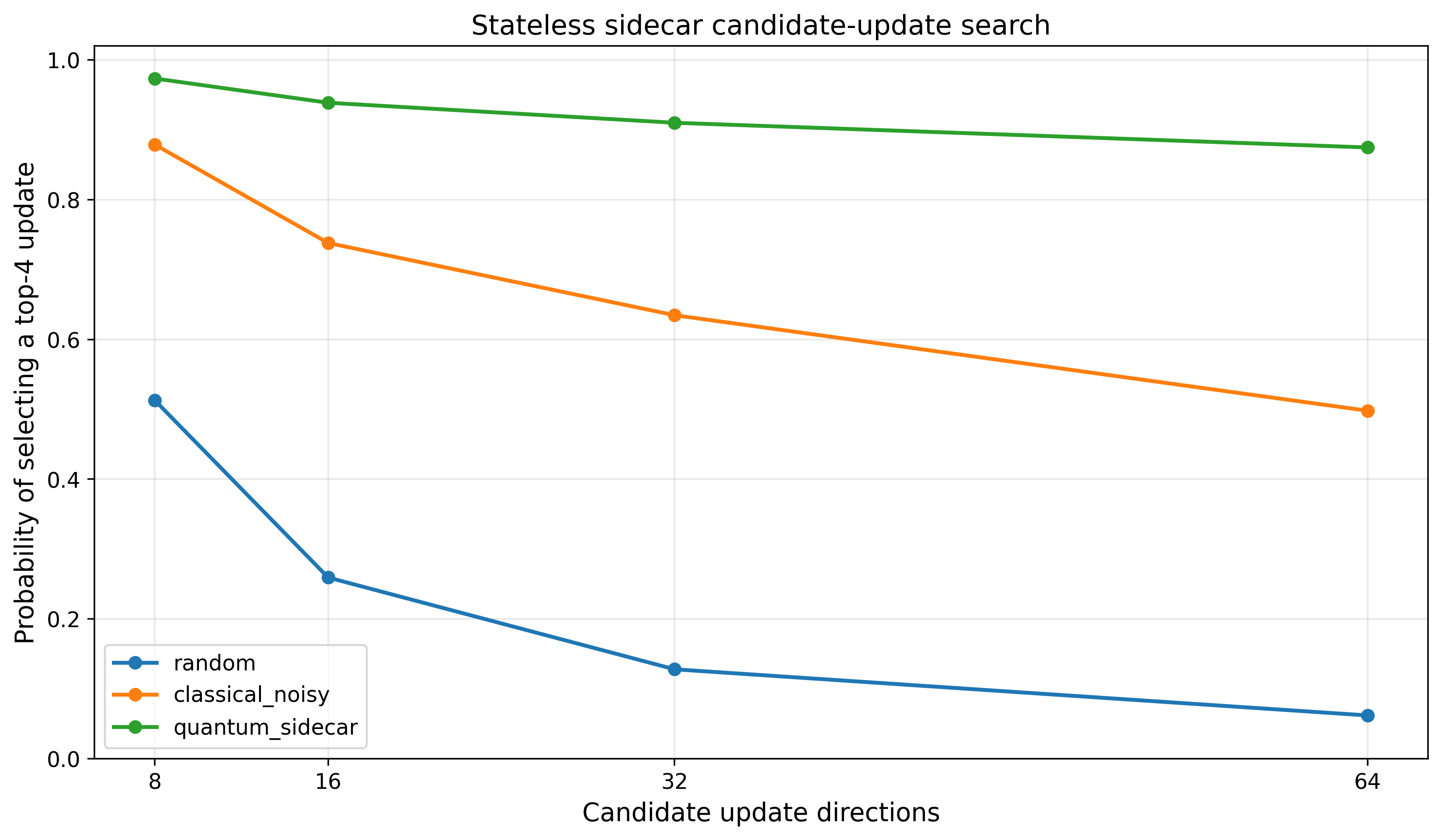}
  \caption{Stateless sidecar candidate-update search. The sidecar-style sampler
  selects top-4 update directions more often than random or noisy classical
  baselines in this synthetic landscape.}
  \label{fig:update_success}
\end{figure}

\begin{figure}[!htbp]
  \centering
  \includegraphics[width=0.88\linewidth]{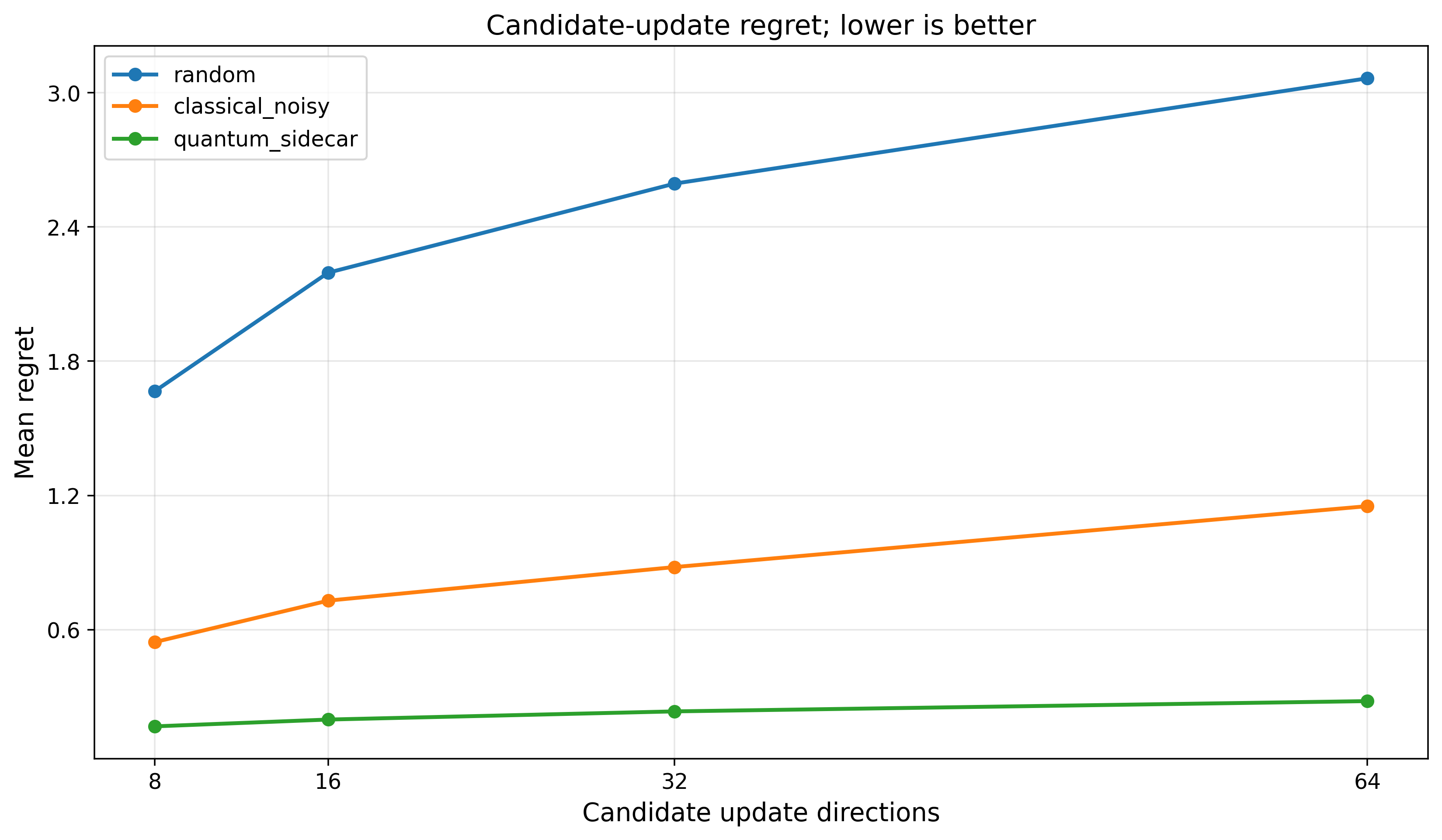}
  \caption{Mean regret for candidate update-direction selection. Lower is
  better.}
  \label{fig:update_regret}
\end{figure}

The stateless training-side experiment models a sidecar that proposes update
directions rather than storing or generating full weights. Each trial creates
$K$ candidate update directions with hidden utility scores. We compare three
selection rules:
\begin{enumerate}
  \item uniform random sampling;
  \item noisy classical softmax selection;
  \item amplitude-biased sidecar-style sampling.
\end{enumerate}
This first experiment is an abstract sidecar sampling model, not a gate-level
quantum circuit simulation and not a claim of guaranteed optimal selection. The
sidecar distribution is sharper and less noisy, representing the hoped-for
effect of a useful quantum proposal module. It is therefore useful mainly as an
interface baseline. Fig.~\ref{fig:update_success} shows the probability of
selecting a top-4 update direction. Fig.~\ref{fig:update_regret} shows mean
regret.

\subsection{Minimal circuit interpretation}

The amplitude-biased sampler can be interpreted as an abstracted output
distribution of a parameterized quantum circuit. In the simplest one-qubit
case,
\[
|0\rangle \xrightarrow{R_y(\theta)} \mathrm{measure},
\]
which gives
\[
P(1)=\sin^2(\theta/2), \qquad P(0)=\cos^2(\theta/2).
\]
A classical score, gradient signal, or candidate utility estimate can choose
$\theta$ and thereby bias a binary update decision. For $n$ qubits, a
parameterized unitary $U(\theta)$ induces a distribution over candidate
indices:
\[
P(z;\theta)=|\langle z|U(\theta)|0^n\rangle|^2,\qquad z\in\{0,1\}^n.
\]
The experiments below use this circuit picture only as a physically valid
interpretation of bounded stochastic sidecar signals. They do not claim that
the displayed sampler solves optimization by itself.

\subsection{Circuit-level QAOA-style stateless sampler}

To remove the strongest artificial assumption in the abstract sampler, we add a
small statevector simulation of a real parameterized quantum circuit. Each
candidate update is represented by a bitstring $z\in\{0,1\}^n$, so the candidate
count is $K=2^n$. We define spins $s_i(z)\in\{-1,+1\}$ and a structured
Ising/QUBO-style utility landscape:
\[
U(z)=\sum_i h_i s_i(z)+\sum_{i<j}J_{ij}s_i(z)s_j(z).
\]
Equivalently, the circuit uses the diagonal utility operator
\[
\hat U=\sum_z U(z)|z\rangle\langle z|.
\]
The stateless sidecar starts from a uniform superposition and applies a
one-layer QAOA-style circuit \cite{farhi2014qaoa}:
\[
|0\rangle^{\otimes n}
\xrightarrow{H^{\otimes n}}
\frac{1}{\sqrt{2^n}}\sum_z |z\rangle,
\]
\[
|\psi(\gamma,\beta)\rangle =
\left(\prod_i e^{-i\beta X_i}\right)
e^{i\gamma \hat U}
\frac{1}{\sqrt{2^n}}\sum_z |z\rangle .
\]
Measuring $|\psi(\gamma,\beta)\rangle$ produces a physically valid distribution
over candidate updates:
\[
P(z;\gamma,\beta)=|\langle z|\psi(\gamma,\beta)\rangle|^2.
\]
This remains a small classical statevector simulation, not a hardware result
and not a quantum-advantage claim. Its purpose is narrower: replacing the
purely assumed amplitude-biased sampler with an explicit circuit family whose
measurement probabilities can be computed, sampled, and compared. Because the
utility is a structured Ising/QUBO Hamiltonian, the phase operator represents a
local $Z$ and $ZZ$ interaction model rather than an arbitrary table lookup. The
experiment does not benchmark the cost of compiling arbitrary black-box
utilities into quantum circuits.

\begin{figure}[!htbp]
  \centering
  \includegraphics[width=0.88\linewidth]{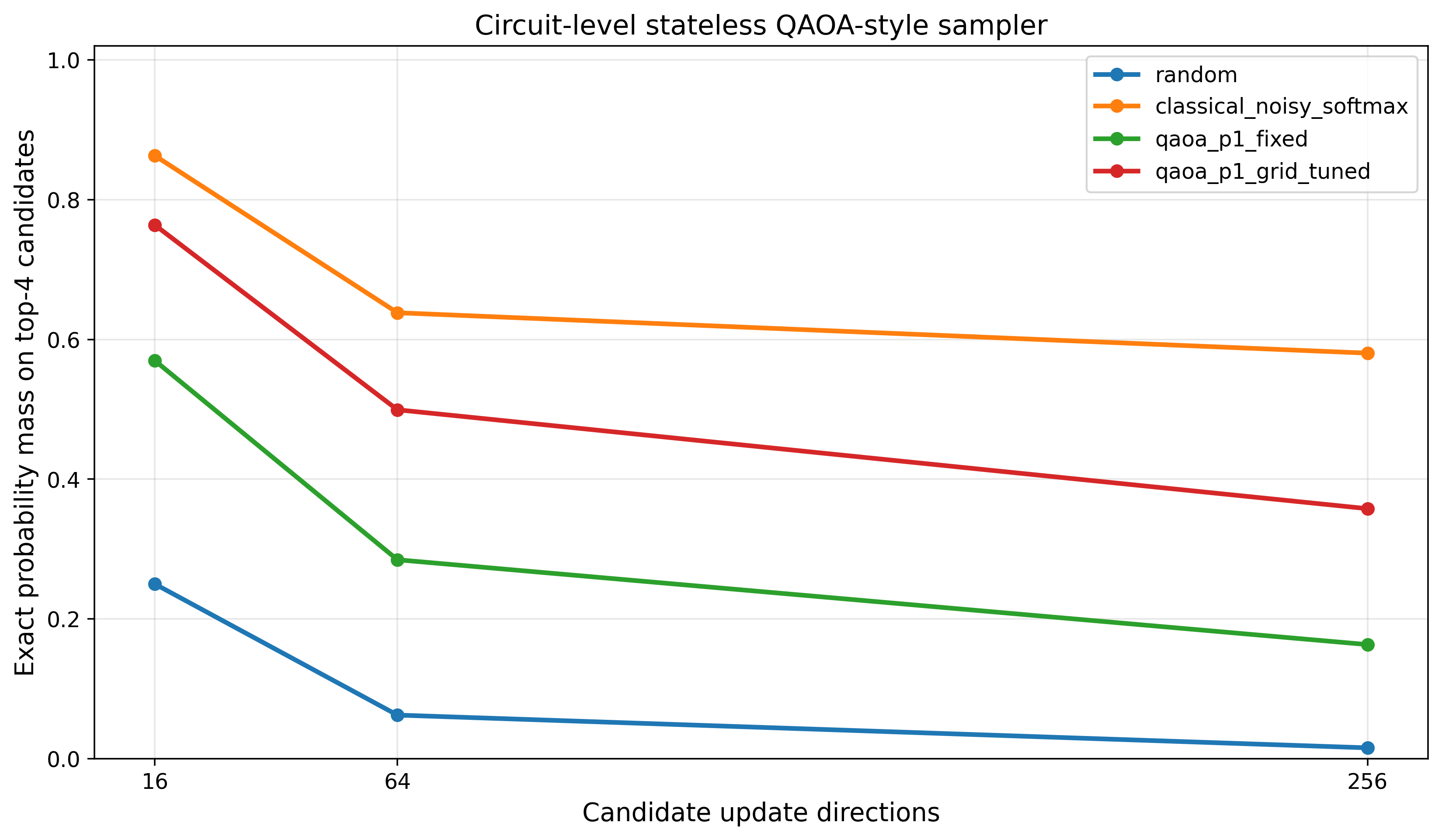}
  \caption{Circuit-level stateless QAOA-style sampler. Unlike the abstract
  sidecar baseline, this distribution is generated by an explicit
  parameterized quantum circuit simulated with a statevector.}
  \label{fig:qaoa_top4}
\end{figure}

\begin{figure}[!htbp]
  \centering
  \includegraphics[width=0.88\linewidth]{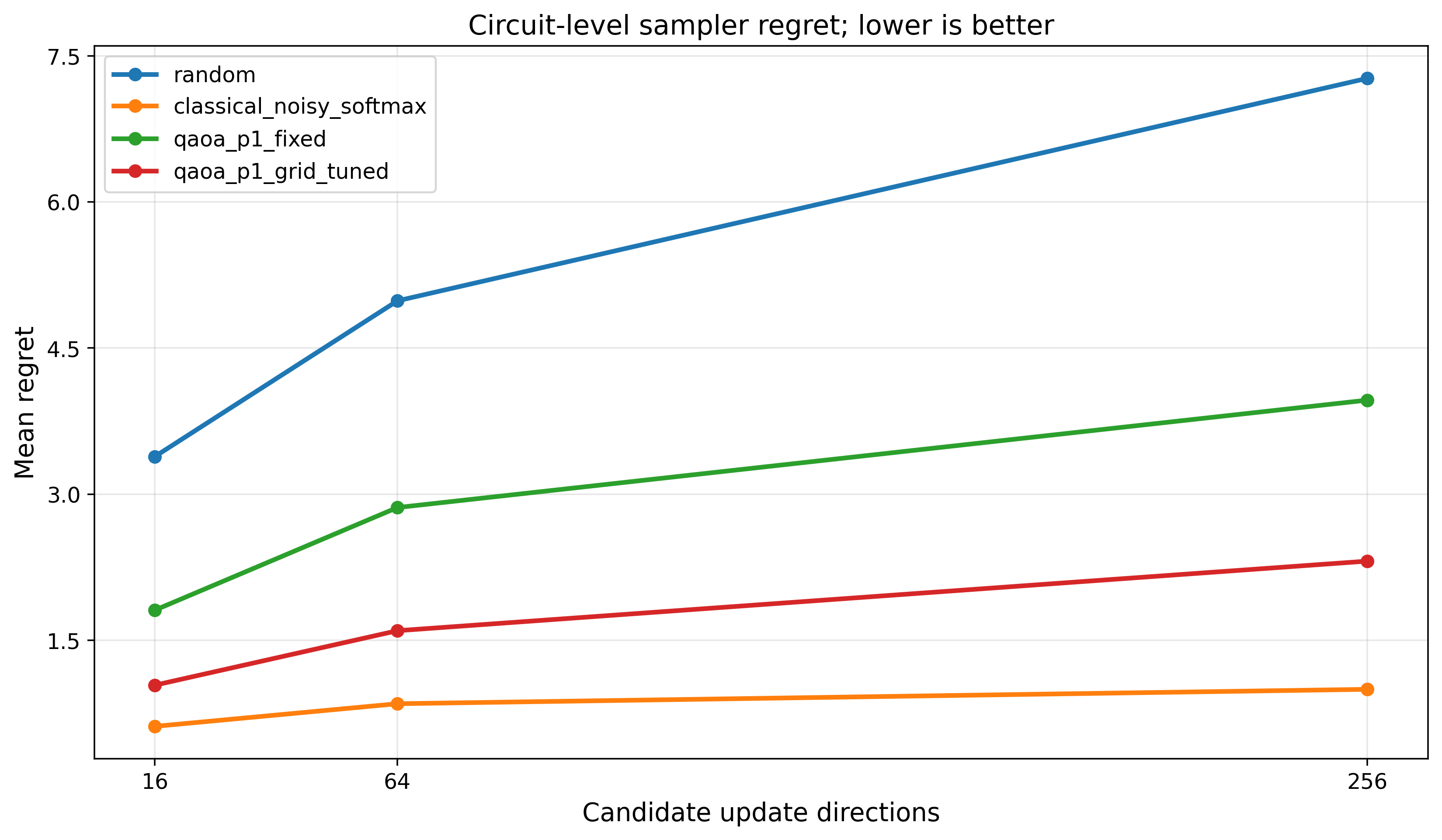}
  \caption{Mean regret for the circuit-level stateless sampler. The
  grid-tuned curve is a diagnostic upper bound for the circuit family, not a
  deployment claim.}
  \label{fig:qaoa_regret}
\end{figure}

\begin{figure}[!htbp]
  \centering
  \includegraphics[width=0.88\linewidth]{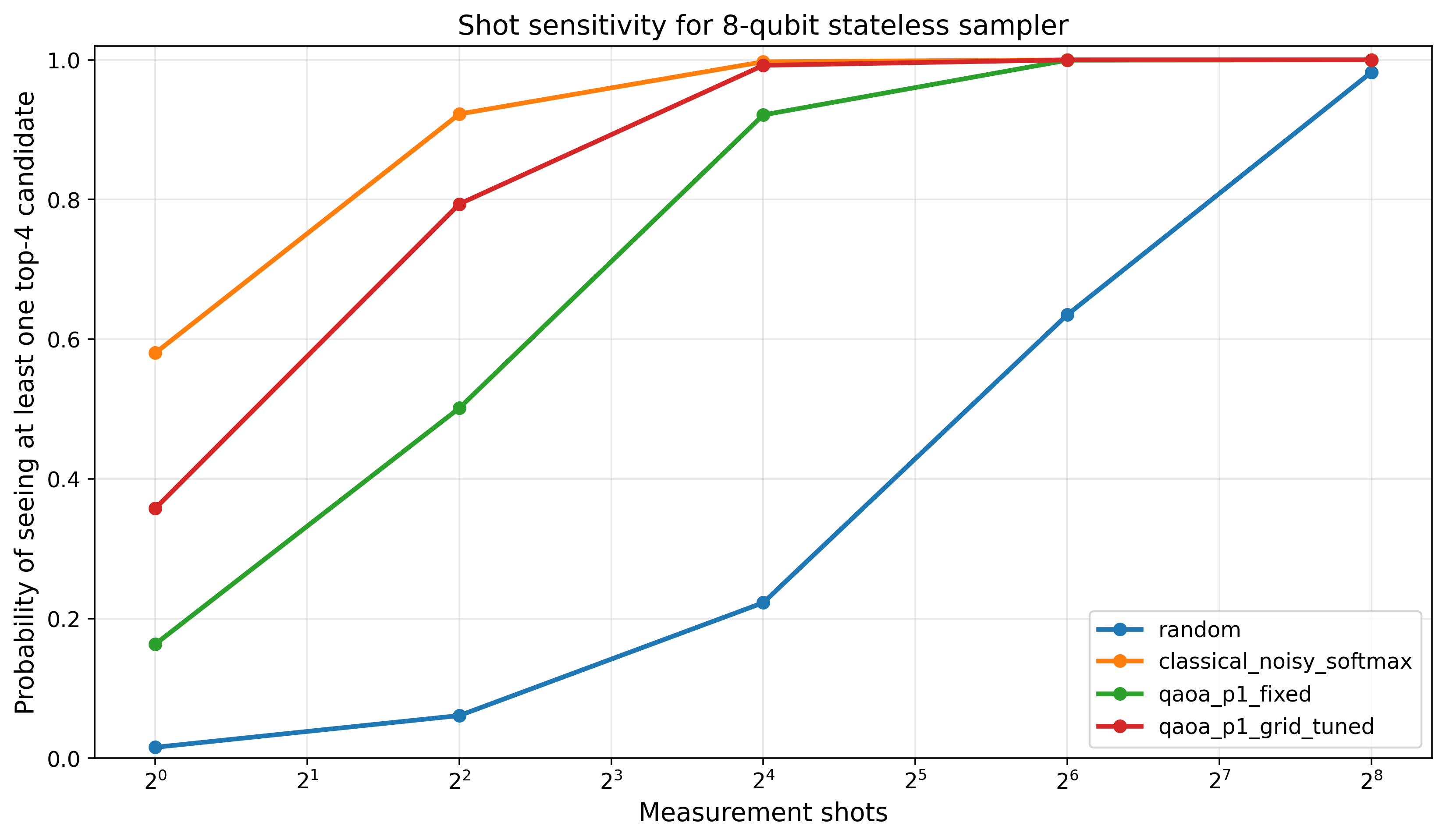}
  \caption{Shot sensitivity for the 8-qubit circuit-level sampler. More
  measurement shots increase the probability of seeing at least one high-utility
  candidate, but they also increase wall-clock cost.}
  \label{fig:qaoa_shots}
\end{figure}

We test $n=4,6,8$ qubits, corresponding to $K=16,64,256$ candidate updates,
over 50 random structured landscapes. We compare uniform random sampling, a
noisy classical softmax sampler, a fixed-parameter one-layer QAOA sampler, and
a grid-tuned one-layer QAOA sampler. The grid-tuned result should be read as an
upper-bound diagnostic for this circuit family, because tuning itself costs
classical computation. Fig.~\ref{fig:qaoa_top4} and Fig.~\ref{fig:qaoa_regret}
show top-4 probability mass and mean regret. Fig.~\ref{fig:qaoa_shots} shows
how repeated measurement shots increase the chance of observing at least one
top-4 candidate in the 8-qubit case.

In these simulations, the QAOA-style circuit places more probability mass on
top candidates than uniform random sampling, but remains below the noisy
classical softmax baseline that receives explicit utility estimates. This is
the intended interpretation: the circuit produces a real non-uniform quantum
measurement distribution, while the experiment does not claim superiority over
well-informed classical sampling.

\FloatBarrier

\subsection{Reset overhead sensitivity}

\begin{figure}[!htbp]
  \centering
  \includegraphics[width=0.88\linewidth]{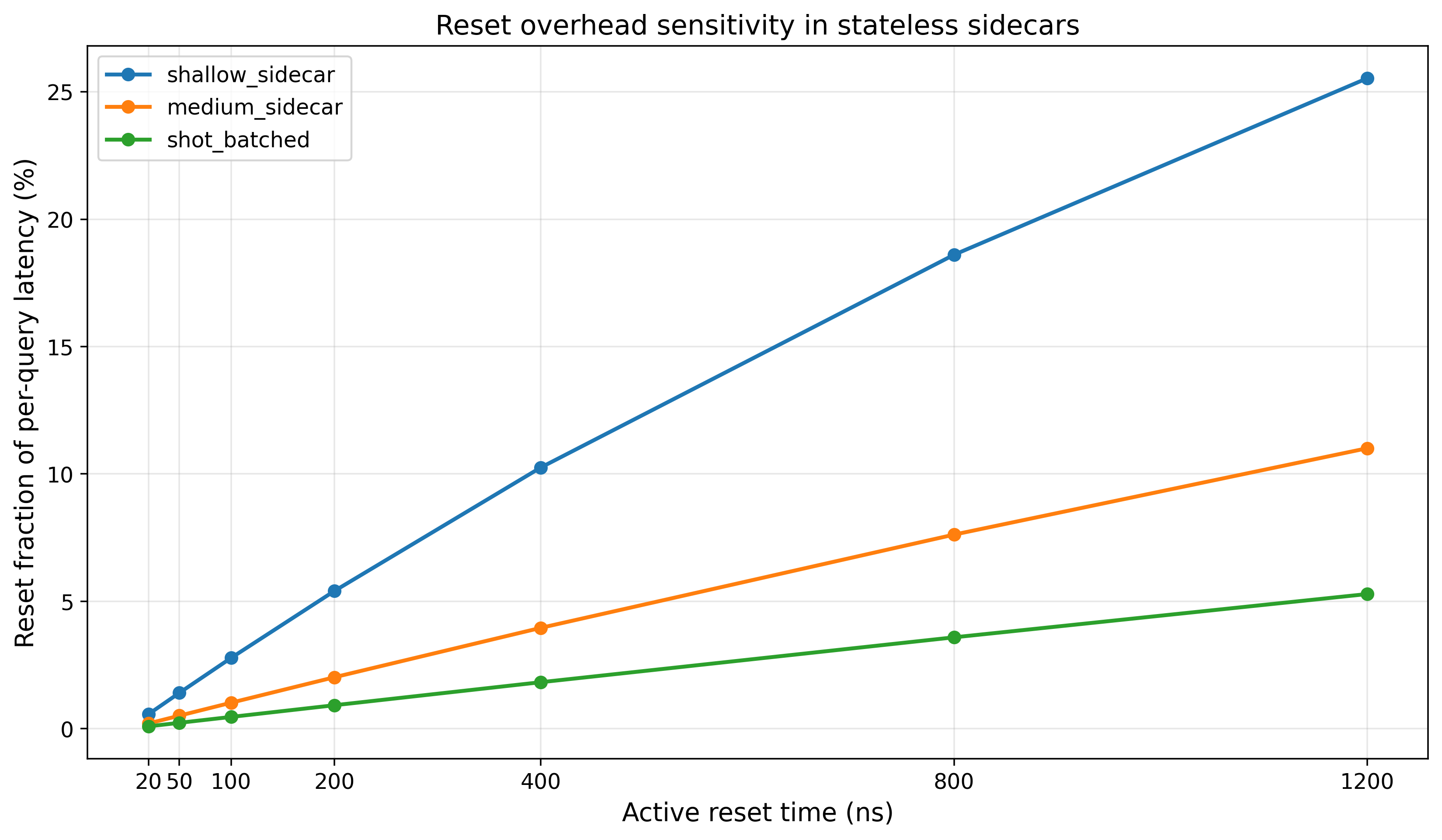}
  \caption{Reset overhead sensitivity in stateless reset-and-reprepare
  sidecars. Reset is not the dominant term in the modeled medium and batched
  scenarios, but it is not the only engineering cost.}
  \label{fig:reset_fraction}
\end{figure}

In a stateless sidecar, all qubits are reset after each query. Superconducting
platforms have demonstrated active reset protocols ranging from feedback-driven
and driven reset to unconditional microwave reset
\cite{geerlings2013driven,magnard2018fast,mckay2018reset}. We therefore model
per-query latency as:
\[
T_{\mathrm{query}} =
T_{\mathrm{prep}}+T_{\mathrm{gate}}+T_{\mathrm{meas}}+
T_{\mathrm{reset}}+T_{\mathrm{classical}}.
\]
Fig.~\ref{fig:reset_fraction} sweeps active reset time from 20 ns to 1200 ns
under three sidecar scenarios. Reset can be a small fraction of query latency,
but this does not eliminate other costs: state preparation, measurement,
shots, circuit depth, and classical-quantum orchestration remain central.

\begin{figure}[!htbp]
  \centering
  \includegraphics[width=0.88\linewidth]{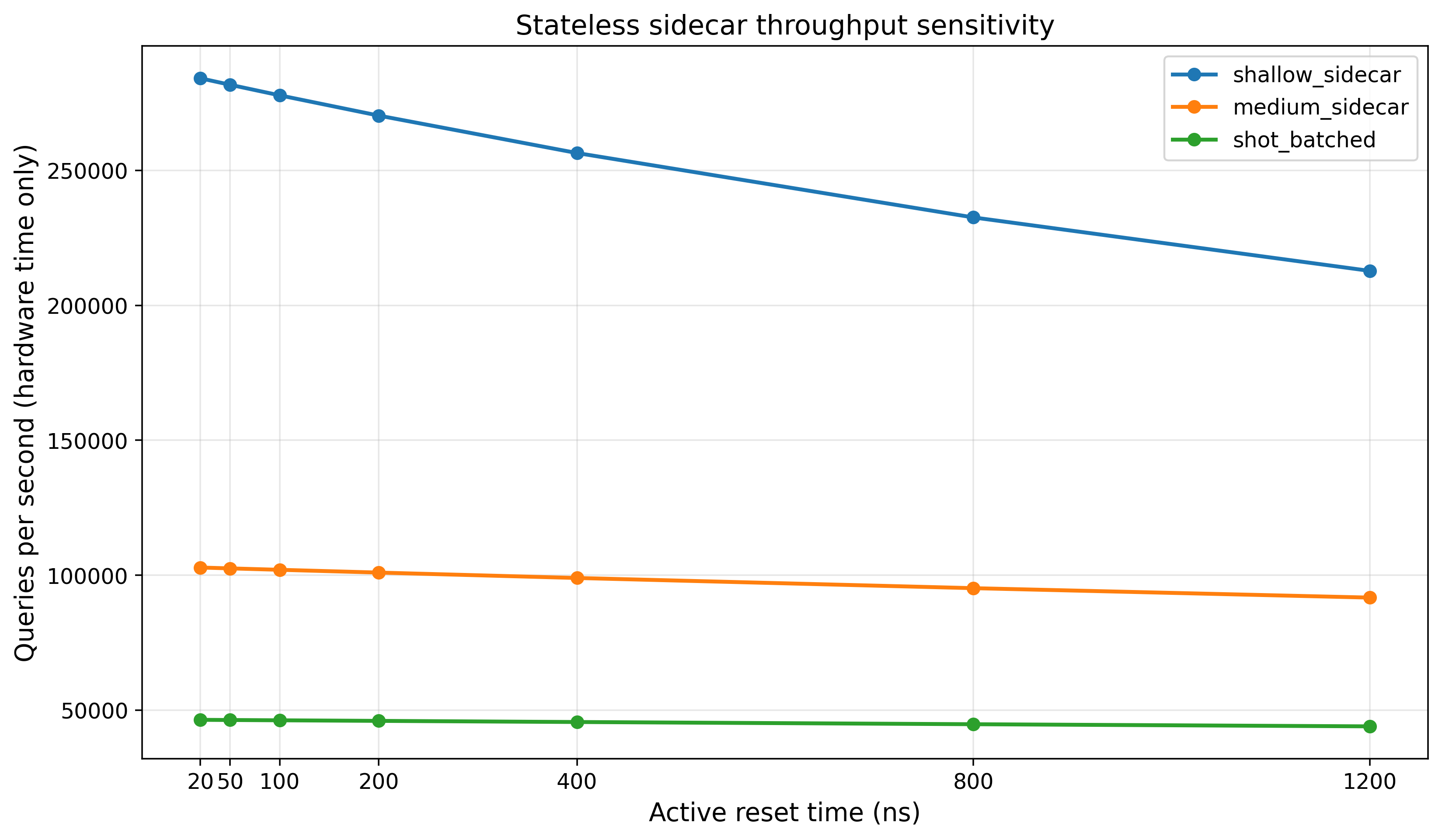}
  \caption{Modeled hardware-time throughput for stateless sidecar queries as a
  function of reset time.}
  \label{fig:reset_throughput}
\end{figure}

\FloatBarrier

\section{Training and Inference Interfaces}

\subsection{Training loop interface}

The stateless update-search experiment suggests a training-side interface:
\[
\mathrm{loss/gradient\ signal}\rightarrow
\mathrm{candidate\ subspace}\rightarrow
\mathrm{sidecar\ proposal}\rightarrow
\mathrm{classical\ evaluation}\rightarrow
W_{t+1}.
\]
The full model remains classical. The sidecar proposes bounded updates that a
classical optimizer accepts, rejects, or reweights.

\subsection{Gradient and update-estimation interfaces}

A sidecar that contains trainable circuit parameters still needs an update
mechanism. This paper does not demonstrate end-to-end LLM training, but three
standard interfaces are compatible with the architecture.

First, parameter-shift rules estimate derivatives of circuit expectation
values for gates generated by Pauli operators with eigenvalues $\pm1$. For a
suitable observable $O$ and a gate of the form $e^{-i\theta P/2}$ with
$P^2=I$, a common two-point form is
\[
\frac{\partial}{\partial\theta}\langle O\rangle_{\theta}
=\frac{1}{2}
\left(\langle O\rangle_{\theta+\pi/2}
-\langle O\rangle_{\theta-\pi/2}\right).
\]
This gives a hardware-measurable gradient for circuit parameters, but the
number of shifted evaluations can scale with the number of trainable
parameters.

Second, stochastic finite-difference methods such as SPSA perturb many
parameters at once and can be useful under shot noise and hardware noise. They
reduce the number of circuit settings per update, but their variance and tuning
cost must be counted.

Third, a stateless sampler can avoid direct gradient estimation and instead
propose candidate updates or control variables. The QAOA-style experiment in
this paper tests this third interface: it evaluates whether an explicit
parameterized circuit can concentrate measurement probability on high-utility
candidate updates in a small structured landscape. Classical validation remains
responsible for accepting, rejecting, or reweighting those proposals.

\subsection{Inference-side consumption}

Inference use is analogous but should not be treated as proof of quantum
advantage. A measured sidecar signal can be consumed as a routing, retrieval,
verification, or decoding-control hint:
\[
s \in \{\mathrm{expert},\mathrm{passage},\mathrm{verifier},\mathrm{path}\}.
\]
The main claim is that such signals define a bounded interface between quantum
sampling and classical inference. A purely classical toy routing illustration
is included only in the appendix to show one possible software consumption
pattern.

\section{Quantum Weight-State Sidecar Outlook}

A more aggressive long-term direction is a quantum weight-state sidecar. This
does not mean storing an entire Transformer weight tensor in amplitudes.
Instead, it means using a trainable quantum state or parameterized quantum
dynamics to represent useful distributions over model-control variables:
adapter updates, expert gates, retrieval policies, verifier choices,
reasoning-path proposals, or decoding controls.

In this setting, ``transient generation'' means generating bounded candidate
updates or control states, not a complete trained model. ``Parallel reasoning''
means amplitude-biased candidate-path proposal followed by classical
verification, not a guarantee that measurement returns the optimal answer.
This framing preserves the quantum-native AI vision while keeping the claim
inside known quantum-information constraints.

\section{Limitations}

This work is intentionally bounded:
\begin{itemize}
  \item It does not clone arbitrary unknown quantum states.
  \item It does not encode full Transformer weights in a small quantum state.
  \item It does not show one-shot collapse into a fully trained model.
  \item It does not guarantee optimal answers from measurement.
  \item The routing interface is a classical appendix illustration, not part of
  the quantum evidence chain.
  \item The update-search experiment is synthetic and does not prove quantum
  advantage.
  \item The circuit-level stateless sampler is a small statevector simulation,
  not a hardware demonstration.
  \item Reset overhead is modeled, not measured on hardware in this work.
\end{itemize}

\section{Conclusion}

We presented a unified quantum sidecar architecture family for hybrid AI
training and inference. The stateful mode uses protected registers and
QND-style readout to preserve selected observables across repeated use. The
stateless mode follows the near-term hardware loop of prepare, evolve, measure,
reset, and repeat, reusing VQA-like circuits as bounded signal sources rather
than presenting a new quantum primitive. Together, these modes support a roadmap
from physically grounded readout and sampling primitives to practical AI
sidecar interfaces for candidate updates, retrieval, routing, and
reasoning-path proposal. The long-term quantum weight-state sidecar vision is
to represent bounded model-control distributions in quantum dynamics, not to
directly replace all explicit classical weights today.

\appendix

\section{Classical Consumption Interface Illustration}

The following toy routing example is intentionally separated from the main
evidence chain. It is a classical software-consumption illustration, not a
quantum simulation, not a new routing algorithm, and not a quantum-advantage
demonstration. Its only purpose is to show how a bounded sidecar-like prior
could enter a Transformer/MoE-style router once some upstream module has
produced a candidate signal.

\begin{figure}[!htbp]
  \centering
  \includegraphics[width=0.88\linewidth]{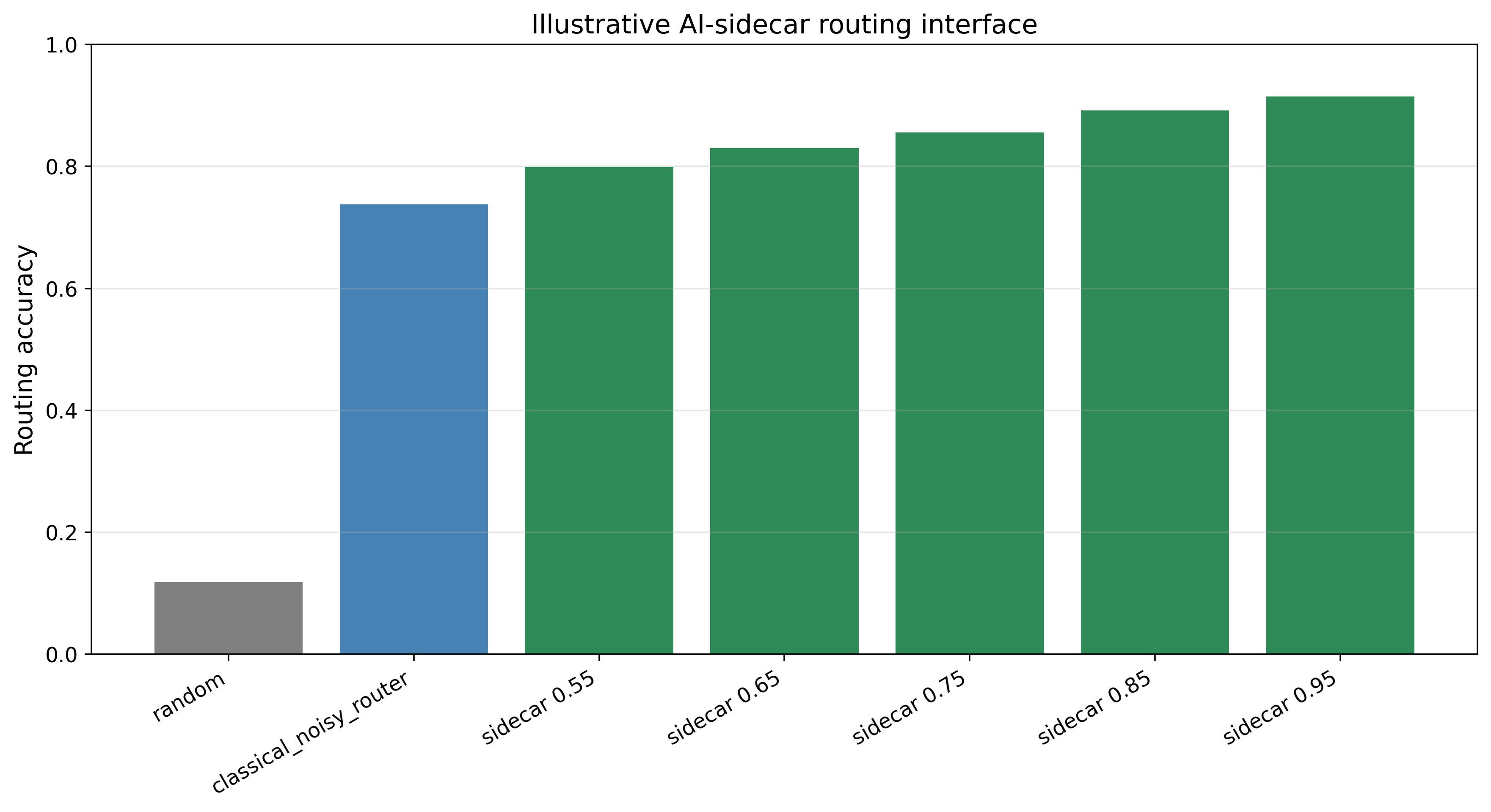}
  \caption{Classical consumption interface illustration for an AI-sidecar
  router. This appendix figure demonstrates a possible software interface only;
  it is not a quantum simulation or a quantum-advantage claim.}
  \label{fig:routing}
\end{figure}

We generate a synthetic 8-expert routing problem and compare a random router, a
noisy classical router, and a sidecar prior with reliability between $0.55$ and
$0.95$. The prior improves accuracy when it is more reliable than random, as
shown in Fig.~\ref{fig:routing}. This result is not used as evidence for the
quantum sidecar primitives studied in the main text.

\FloatBarrier

\section*{Reproducibility}

All figures and CSV files are generated by:
\begin{verbatim}
python simulate_sidecar_architectures.py
\end{verbatim}


\begin{thebibliography}{99}

\bibitem{vaswani2017attention}
A. Vaswani et al.,
``Attention is all you need,''
\emph{Advances in Neural Information Processing Systems}, 2017.

\bibitem{brown2020language}
T. Brown et al.,
``Language models are few-shot learners,''
\emph{Advances in Neural Information Processing Systems}, 2020.

\bibitem{fedus2022switch}
W. Fedus, B. Zoph, and N. Shazeer,
``Switch Transformers: Scaling to trillion parameter models with simple and
efficient sparsity,''
\emph{Journal of Machine Learning Research}, vol. 23, no. 120, pp. 1--39, 2022.

\bibitem{lewis2020retrieval}
P. Lewis et al.,
``Retrieval-augmented generation for knowledge-intensive NLP tasks,''
\emph{Advances in Neural Information Processing Systems}, 2020.

\bibitem{wootters1982single}
W. K. Wootters and W. H. Zurek,
``A single quantum cannot be cloned,''
\emph{Nature}, vol. 299, pp. 802--803, 1982.

\bibitem{dieks1982communication}
D. Dieks,
``Communication by EPR devices,''
\emph{Physics Letters A}, vol. 92, no. 6, pp. 271--272, 1982.

\bibitem{nielsen2010quantum}
M. A. Nielsen and I. L. Chuang,
\emph{Quantum Computation and Quantum Information},
10th anniversary ed., Cambridge University Press, 2010.

\bibitem{biamonte2017quantum}
J. Biamonte et al.,
``Quantum machine learning,''
\emph{Nature}, vol. 549, pp. 195--202, 2017.

\bibitem{schuld2018supervised}
M. Schuld and F. Petruccione,
\emph{Supervised Learning with Quantum Computers},
Springer, 2018.

\bibitem{havlicek2019supervised}
V. Havlicek et al.,
``Supervised learning with quantum-enhanced feature spaces,''
\emph{Nature}, vol. 567, pp. 209--212, 2019.

\bibitem{peruzzo2014variational}
A. Peruzzo et al.,
``A variational eigenvalue solver on a photonic quantum processor,''
\emph{Nature Communications}, vol. 5, p. 4213, 2014.

\bibitem{cerezo2021variational}
M. Cerezo et al.,
``Variational quantum algorithms,''
\emph{Nature Reviews Physics}, vol. 3, pp. 625--644, 2021.

\bibitem{lloyd2018quantum}
S. Lloyd and C. Weedbrook,
``Quantum generative adversarial learning,''
\emph{Physical Review Letters}, vol. 121, p. 040502, 2018.

\bibitem{dallaire2018quantum}
P.-L. Dallaire-Demers and N. Killoran,
``Quantum generative adversarial networks,''
\emph{Physical Review A}, vol. 98, p. 012324, 2018.

\bibitem{benedetti2019generative}
M. Benedetti et al.,
``A generative modeling approach for benchmarking and training shallow quantum
circuits,''
\emph{npj Quantum Information}, vol. 5, p. 45, 2019.

\bibitem{mitarai2018quantum}
K. Mitarai, M. Negoro, M. Kitagawa, and K. Fujii,
``Quantum circuit learning,''
\emph{Physical Review A}, vol. 98, p. 032309, 2018.

\bibitem{schuld2019evaluating}
M. Schuld, V. Bergholm, C. Gogolin, J. Izaac, and N. Killoran,
``Evaluating analytic gradients on quantum hardware,''
\emph{Physical Review A}, vol. 99, p. 032331, 2019.

\bibitem{spall1992spsa}
J. C. Spall,
``Multivariate stochastic approximation using a simultaneous perturbation
gradient approximation,''
\emph{IEEE Transactions on Automatic Control}, vol. 37, no. 3, pp. 332--341,
1992.

\bibitem{braginsky1996qnd}
V. B. Braginsky and F. Y. Khalili,
\emph{Quantum Measurement},
Cambridge University Press, 1996.

\bibitem{gottesman1997stabilizer}
D. Gottesman,
``Stabilizer codes and quantum error correction,''
Ph.D. dissertation, California Institute of Technology, 1997.

\bibitem{qiskit2024}
Qiskit contributors,
``Qiskit: An open-source framework for quantum computing,''
\url{https://www.ibm.com/quantum/qiskit}, 2024.

\bibitem{farhi2014qaoa}
E. Farhi, J. Goldstone, and S. Gutmann,
``A quantum approximate optimization algorithm,''
\emph{arXiv preprint arXiv:1411.4028}, 2014.

\bibitem{geerlings2013driven}
K. Geerlings et al.,
``Demonstrating a driven reset protocol for a superconducting qubit,''
\emph{Physical Review Letters}, vol. 110, p. 120501, 2013.

\bibitem{magnard2018fast}
P. Magnard et al.,
``Fast and unconditional all-microwave reset of a superconducting qubit,''
\emph{Physical Review Letters}, vol. 121, p. 060502, 2018.

\bibitem{mckay2018reset}
D. C. McKay et al.,
``A pulsed reset protocol for fixed-frequency superconducting qubits,''
\emph{arXiv preprint arXiv:1802.08980}, 2018.

\end{thebibliography}
\end{document}